\documentstyle[12pt,aaspp4,epsf]{article}

\def\fo{\hbox{{1}\kern-.25em\hbox{l}}}
\def\fnote#1#2{\begingroup\def\thefootnote{#1}\footnote{#2}\addtocounter
{footnote}{-1}\endgroup}

\renewcommand{\thefootnote}{\fnsymbol{footnote}}

\def\beq{\begin{equation}}
\def\eeq{\end{equation}}
\def\eq{\end{equation}}
\def\to{\rightarrow}

\def\bsg{\ifmmode B\to X_s\gamma\else $B\to X_s\gamma$\fi}
\def\bsll{\ifmmode B\to X_s\ell^+\ell^-\else $B\to X_s\ell^+\ell^-$\fi}
\def\bstt{\ifmmode B\to X_s\tau^+\tau^-\else $B\to X_s\tau^+\tau^-$\fi}
\def\shat{\ifmmode \hat{s}\else $\hat{s}$\fi}

\newcommand{\newc}{\newcommand}

\newc{\lcal}{\int {\cal L}dt}
 
\newc{\LSP}{{\chi^0_1}}
\newc{\stauR}{{\tilde \tau_R}}
\newc{\stau}{{\tilde \tau_1}}
\newc{\mstop}{m_{\tilde{t}}}
\newc{\mHpm}{m_{H^\pm}}
\newc{\gsim}{\lower.7ex\hbox{$\;\stackrel{\textstyle>}{\sim}\;$}}
\newc{\lsim}{\lower.7ex\hbox{$\;\stackrel{\textstyle<}{\sim}\;$}}
\newc{\ie}{{\it i.e.}}          
\newc{\eg}{{\it e.g.}}          
\newc{\kev}{\hbox{\rm\,keV}}            
\newc{\mev}{\hbox{\rm\,MeV}}            
\newc{\gev}{\hbox{\rm\,GeV}}            
\newc{\tev}{\hbox{\rm\,TeV}}
\newc{\xpb}{\hbox{\rm\, pb}}
\newc{\xfb}{\hbox{\rm\, fb}}

\newc{\mtop}{m_t}
\newc{\mbot}{m_b}
\newc{\mz}{m_Z}
\newc{\mw}{M_W}
\newc{\alphasmz}{\alpha_s(m_Z^2)}
\newc{\swsq}{\sin^2\theta_W}
\newc{\tw}{\tan\theta_W}
\newc{\cw}{\cos\theta_W}
\newc{\sw}{\sin\theta_W}
\newc{\BR}{\hbox{\rm BR}}
\newc{\zbb}{Z\to b\bar}
\newc{\Gb}{\Gamma (Z\to b\bar b)}
\newc{\Gh}{\Gamma (Z\to \hbox{\rm hadrons})}
\newc{\rbsm}{R_b^\hbox{\rm sm}}
\newc{\rbsusy}{R_b^\hbox{\rm susy}}
\newc{\drb}{\delta R_b}

\newc{\sgn}{\mbox{sgn}}
\newc{\tbeta}{\tan\beta}
\newc{\uL}{{\tilde u_L}}
\newc{\uR}{{\tilde u_R}}
\newc{\cL}{{\tilde c_L}}
\newc{\cR}{{\tilde c_R}}
\newc{\tL}{{\tilde t_L}}
\newc{\tR}{{\tilde t_R}}
\newc{\dL}{{\tilde d_L}}
\newc{\dR}{{\tilde d_R}}
\newc{\sL}{{\tilde s_L}}
\newc{\sR}{{\tilde s_R}}
\newc{\bL}{{\tilde b_L}}
\newc{\bR}{{\tilde b_R}}
\newc{\eL}{{\tilde e_L}}
\newc{\eR}{{\tilde e_R}}
\newc{\mhp}{m_{H^\pm}}
\newc{\mhalf}{m_{1/2}}
\newc{\emt}{{e/\mu /\tau}}

\newc{\lR}{\tilde{l}_R}
\newc{\lL}{\tilde{l}_L}
\newc{\nL}{\tilde{\nu}_L}
\newc{\na}{\chi^0_1}
\newc{\nb}{\chi^0_2}
\newc{\nc}{\chi^0_3}
\newc{\nd}{\chi^0_4}
\newc{\ca}{\chi^{\pm}_1}
\newc{\cb}{\chi^{\pm}_2}
\newc{\camp}{\chi^\mp_1}
\newc{\cbmp}{\chi^\mp_1}
\newc{\capos}{\chi^{+}_1}
\newc{\caneg}{\chi^{-}_1}
\newc{\phit}{\phi_t}
\newc{\phib}{\phi_b}
\newc{\phiew}{\phi_{ew}}
\newc{\htz}{h^0_t}
\newc{\hbz}{h^0_b}
\newc{\hewz}{h^0_{ew}}
\newc{\hsmz}{h^0_{sm}}
\newc{\huz}{h^0_u}
\newc{\hsusyz}{h^0_{susy}}

\def\PLB{Phys.\ Lett. B}
\def\PRD{Phys.\ Rev.\ D}

\def\APJ{ApJ}
\def\beq{\begin{equation}}
\def\eeq{\end{equation}}
\def\bea{\begin{eqnarray}}
\def\eea{\end{eqnarray}}
\def\slashchar#1{\setbox0=\hbox{$#1$}           
   \dimen0=\wd0                                 
   \setbox1=\hbox{/} \dimen1=\wd1               
   \ifdim\dimen0>\dimen1                        
      \rlap{\hbox to \dimen0{\hfil/\hfil}}      
      #1                                        
   \else                                        
      \rlap{\hbox to \dimen1{\hfil$#1$\hfil}}   
      /                                         
   \fi}                                         %
\catcode`@=11
\long\def\@caption#1[#2]#3{\par\addcontentsline{\csname
  ext@#1\endcsname}{#1}{\protect\numberline{\csname
  the#1\endcsname}{\ignorespaces #2}}\begingroup
    \small
    \@parboxrestore
    \@makecaption{\csname fnum@#1\endcsname}{\ignorespaces #3}\par
  \endgroup}
\catcode`@=12

\begin{document}

\vspace{-2.0truecm}
\begin{flushright}
SLAC-PUB-7806 \\
CERN-TH/98-362 \\
hep-ph/9811325 \\
November 1998 \\
\end{flushright}
\vspace{-0.5truecm}

\title{Illuminating dark matter and primordial black holes \\
with interstellar antiprotons}

\author{
   James D.~Wells\altaffilmark{1,2},
   Alexander~Moiseev\altaffilmark{3},
   Jonathan F.~Ormes\altaffilmark{3,4}
}

\altaffiltext{0}{${}^1$CERN, Theory Division, CH-1211 Geneva 23, Switzerland}
\altaffiltext{0}{${}^2$Stanford Linear Accelerator Center, Stanford
  University, Stanford, CA 94309}
\altaffiltext{0}{${}^3$NASA/Goddard Space Flight Center, Greenbelt, MD 20771}
\altaffiltext{0}{${}^4$This work was started while on 
  leave at Stanford Linear Accelerator Center}
\begin{abstract}

Interstellar antiproton fluxes can arise from dark matter
annihilating or decaying into quarks or gluons that subsequently
fragment into antiprotons.  Evaporation of primordial black holes also
can produce a significant antiproton cosmic-ray flux.  Since the
background of secondary antiprotons from spallation has an interstellar
energy spectrum that peaks at $\sim 2\gev$ and falls rapidly for
energies below this, low-energy measurements of cosmic antiprotons are
useful in the search for exotic antiproton sources.  However,
measurement of the flux near the earth is challenged by significant
uncertainties from the effects of the solar wind.  We suggest evading
this problem and more effectively probing dark-matter signals by
placing an antiproton spectrometer aboard an interstellar probe
currently under discussion.  We address the experimental challenges of
a light, low-power-consuming detector, and present an initial design
of such an instrument.  This experimental effort could significantly
increase our ability to detect, and have confidence in, a signal of
exotic, nonstandard antiproton sources.  Furthermore, solar modulation
effects in the heliosphere would be better quantified and understood
by comparing results to inverse modulated data derived from existing
balloon and space-based detectors near the earth.

\end{abstract}

\clearpage
\section{Introduction}
\bigskip

Experimental and theoretical investigations of galaxy rotation,
large-scale structure formation, big bang nucleosynthesis, and other
observables appear to have converged on the necessity of dark 
matter (Trimble 1987; Sikivie 1995).
Dark matter is non-luminous matter which comprises a significant
amount of the mass density of the universe.  It may be mostly in the
form of weakly interacting massive particles (WIMPs), and it may
be due, at least in part, to primordial black holes (PBHs). 
It is not enough to hypothesize new stable
particles and other interesting cosmological
remnants.  These objects must be found or excluded
experimentally to make progress
in our understanding of how the universe is put together.  It is toward
this goal that we write this paper.  

There are numerous possible ways to find dark matter.  
If the WIMPs interact
with ordinary matter then collisions of WIMPs with nuclei in the sun and
earth could dissipate the WIMPs' energy enough to be captured.  When 
they accrue in the sun and earth they will annihilate rapidly.
The annihilation products will then fragment into neutrinos,
some of which have
high momentum directed toward earth-based neutrino telescopes.
Current estimates  in the supersymmetric case suggest
that this is a difficult way to see dark matter unless there is an
enhanced spin-dependent nuclear interaction
(Jungman, Kamionkowski \&\ Griest 1996; Diehl et al. 1995).

One can search for dark matter also in a cryogenic 
detector (Cabrera 1998).
With enough target material, it might be possible to see ambient
WIMP collisions with cooled nuclei in the laboratory.
The current estimate of the mass density of WIMPs in our solar system
is $\rho \simeq 0.3\gev\, /{\rm cm^3}$, which means that there is
only a handful of weak-scale mass WIMPs per cubic meter.  
Finding a signal in cryogenic detectors
would be difficult unless there is a large coherent scalar interaction
between the WIMP and the heavy nuclei in the cryogenic detector.
Scalar interactions couple coherently to the nucleus, and so the
counting rate is proportional to the mass of the nucleus not the mass
of the constituent protons and neutrons.  Therefore, larger
mass nuclei are generally preferred for cryogenic detectors to take
advantage of this possibility (Griest 1988).

Both the neutrino and cryogenic experimental programs rely on the
dark matter interacting with ordinary matter.  However, it is 
possible that the dark matter does not interact directly with ordinary matter.
It is also possible that our local density is much smaller than the
standard estimates indicate.
In these circumstances, the above two experimental approaches
may not find a signal for dark matter.  Additional experimental methods are
necessary to complete the search strategy for dark matter.

Annihilations and decays of dark matter in the galactic halo can yield ordinary
particles even though the dark matter does not interact directly with them.
This is true in the case of dark matter
which carries no standard model quantum numbers.  The annihilation
or decay products are a sum over particles and antiparticles
with zero net charge.  
Supersymmetry dark matter candidates generally have no quantum numbers, and 
fit in this category.  This is also true of
primordial black holes.  Therefore, searches for annihilations or decays
of dark matter
into photons, positrons and antiprotons in cosmic rays are useful probes.

There are several sources of photons from dark matter.  Annihilations
into quarks and gluons which fragment into neutral pions which then 
decay into photons is one source.  In the case of WIMP annihilation
this gives a continuous spectrum of photons steeply falling 
with energy, $0\lsim E_\gamma \lsim m_{\rm WIMP}$.  The 
standard photon background is also
a steeply falling function of $E_\gamma$.  This makes continuum photon
signals difficult to resolve from background.  
However, one can use the measured photon background as a veto against
all combinations 
of particle physics and astrophysics models which yield an unacceptably 
large photon flux.  Another way to perhaps see dark matter is from interactions
with electrons near the accretion disk of an active galaxy 
(Bloom \&\ Wells 1998).  High energy 
photons emitted from an AGN at high angles may be indicative of non-standard
interactions near the AGN.  The non-uniqueness of the photon signal energy
distribution and the expected small rate make this technique challenging
as well.  Yet another source
of photons, and perhaps the most promising, 
is WIMP annihilation directly into two photons.  Since the
dark matter is not charged, this must proceed via a loop diagram.
Since these loop diagrams are suppressed by factors of $\alpha/4\pi$ (where
$\alpha=1/137$ is the QED coupling constant), the total rate for
this observable 
is small.  However, since the expected velocity of the WIMPs is 
non-relativistic ($\sim 10^{-3}c$), the resulting photons are 
monochromatic.  High photon energy
resolution detectors may be able to see this type of signal, but 
the low rate is still a challenge 
(Bergstr\"om, Ullio, \&\ Buckley 1998; Bloom et al. 1998).

Positrons may also 
arise from the annihilation and decay products of dark matter.  
The largest source of positrons is from fragmentation
of quarks into charged pions which then decay to positrons.  The energy
spectrum here is also continuous, and resolving signal positrons from
background positrons is difficult.  However, if the dark matter annihilates
into a $W$ boson (or a top quark which then decays to a $W$ boson) then
the positron from $W^+\to e^+\nu$ may have a non-trivial energy profile
with distinctive
bumps and peaks in the $e^+/(e^++e^-)$ 
spectrum (Turner \&\ Wilczek 1990; Kamionkowski \&\ Turner 1991;
Diehl et al. 1995; Barwick et al. 1997 and 1998).
However, the peakedness of
the signal gets washed out somewhat when the parent particles of the positrons
are boosted by a significant amount (i.e., $m_{\rm WIMP}\gg m_W$).  
Furthermore, the signal is not
present if a $W^+$ is not present in the decay chain.  The positron signal
also suffers from QED energy loss effects which serve to broaden
any positron peak when it is propagated over the galaxy on its way to the
detector.

Each of these modes of detecting dark matter has difficulties
that are hard to control theoretically and experimentally.  Here, we will
be mainly interested in antiproton searches of dark matter.  When dark matter
annihilates or decays it can produce QCD jets which fragment into antiprotons.
If enough antiprotons are produced in this process it is hoped that
they will be detected above background.  
As will be discussed below, antiprotons are qualitatively different
than continuum photons and positrons as probes of dark matter.  The unique
feature of antiproton experiments is the peaking behavior in the background
antiproton energy distribution.  The signal does not have this feature which
makes separation of signal from background possible.

In the next section we review
the standard sources of the antiproton cosmic ray component
and discuss the challenges experiments face when trying to extract a
signal of dark matter annihilations and/or decays.  The challenges
are significant for the experiments being performed near the earth, 
where antiprotons are battered by a hefty solar wind.  
We then discuss the experimental challenges of detecting low-energy
antiprotons at the necessary sensitivities.  The size, weight,
and power consumption requirements for any apparatus bound for 
interstellar space is confronted, an initial design is presented, 
and sensitivities are estimated.  We then make some concluding
comments in the last section.

\section{Antiproton probes of dark matter and PBHs}
\bigskip

It was long ago acknowledged that one could potentially see dark matter
by looking for an excess of antiprotons in cosmic rays from
the annihilation or decay products of dark 
matter particles (Stecker, Rudaz, \&\ Walsh 1985; Rudaz \&\ Stecker 1988;
Ellis et al. 1988; Jungman \&\ Kamionkowski 1994).
This approach
is unique since it is the {background} which has a non-trivial
energy dependence, whereas the signal is a steeply falling function
with increasing energy of the antiproton.  The background antiprotons
are produced by interactions of the high-energy proton cosmic rays
with interstellar gas and dust (mostly other protons essentially at rest).  
To conserve baryon
number $pp$ collisions must produce a final state with baryon number 2.
The lowest multiplicity final state where this is possible, and where
antiprotons are also created is $pp\to p\bar p pp $.
To produce this final state, the incident proton must have kinetic energy
$K>6\gev$ with respect to the at-rest proton target
($K\equiv E_{\bar p}-m_{\bar p}$).
At threshold in the center-of-mass frame, the $\bar p$ is at rest.
Boosting back into the lab (or cosmic) frame, the $\bar p$ has a non-zero
kinetic energy of about $\sim 2\gev$.
To produce a $\bar p$ with kinetic
energy any less than this requires a higher energy proton impinging
on the at-rest proton such that the $\bar p$ in the center-of-mass
frame has momentum in a direction opposite to the boost direction
which takes the center-of-mass frame back to the lab frame.
Since the primary proton flux
is a steeply falling function of energy, the secondary
antiproton flux must decrease for kinetic energies less than $\sim 2\gev$.
In Fig.~1 the solid line is the background interstellar
$\bar p$ flux from secondary processes, including $pp$ collisions
and proton collisions with heavy 
nuclei (Simon, Molnar, \&\ Roesler 1998; See also, 
Stephens \&\ Golden 1987; Webber \&\ Potgieter 1989; Gaisser \&\ Schaefer 1992;
Labrador \&\ Mewaldt 1997; Bottino et al. 1998).

Several attempts have been made to correlate theoretical models with
the antiproton spectrum.
One attractive theoretical framework is
low energy supersymmetry which naturally allows for a stable lightest
supersymmetric partner (LSP).  In supersymmetric theories with gauge
coupling unification and radiative electroweak symmetry breaking
(Higgs mechanism derived radiatively), the
relic abundance of the LSP is typically near closure density, making
the LSP an excellent cold dark matter 
candidate (Drees \&\ Nojiri 1993; Kane et al. 1994; Diehl et al. 1995).
More flexible approaches to supersymmetry also predict the LSP able to have
a significant component of the cold dark matter of the
universe (Wells 1998).
The LSP in these cases is the lightest
neutralino which carries no quantum numbers and can annihilate efficiently
into standard model final states such as $b\bar b$.  The fragmentation
products of these final states include antiprotons, and predictions
can be made for how the LSP annihilations change the antiproton energy
spectrum from the expectations of secondary mechanisms.  

The interstellar $\bar p$ flux from supersymmetry 
is represented (Diehl et al. 1995) by
\beq
\label{susyflux}
\Phi^{\rm IS}_{\bar p}(K) =\frac{(\sigma v)_{AB}}{4\pi}
 \frac{\rho^2_{\rm loc}}{m_\chi^2}(v_{\bar p}\tau_{\bar p})
 F_{\bar p/AB}(K,m_\chi)
\eeq
where $\rho_{\rm loc}\simeq 0.3~\gev /{\rm cm}^3$,
$K$ is the kinetic energy of an antiproton, $v_{\bar p}$ is
the velocity, and $\tau_{\bar p}$ is the containment time in the
galaxy ($\sim 5\times 10^7~{\rm yrs}$).  $F_{\bar p/AB}(K,m_\chi)$
is the fragmentation function quantifying the multiplicity and
energy distribution of antiprotons in
\beq
\chi\chi \to AB \to \bar p + \cdots .
\eeq
Clearly, Eq.~\ref{susyflux} contains significant
particle physics uncertainties in $m_\chi$ and
the annihilation cross-section
$(\sigma v)_{AB}$ from not knowing the underlying supersymmetry breaking
parameters.  However, for any given set of supersymmetric parameters these
values are reliably computed.  There is also some small and insignificant
particle physics 
uncertainties in $F_{\bar p/AB}(K,m_\chi)$, but our knowledge of
the remaining pieces of Eq.~\ref{susyflux} are limited by astrophysics
uncertainties.  Our ignorance of many astrophysics parameters 
constitutes at least a factor of $10$ uncertainty
in the signal flux predictions by themselves.

The only energy range where the antiprotons from LSP annihilation are expected
to be more copious than antiprotons from spallation is in the sub-GeV
region.  Unfortunately, 
this is precisely the area where solar modulation is most volatile.
Solar modulation not only attenuates the flux of
low energy protons and antiprotons, it also can dissipate some of the
energy of these particles as they travel within its domain, which 
extends $\lsim 60$~AU (Simpson 1989) radially from the sun.
(For an illustration of solar wind effects on low energy
protons and antiprotons, see Figs.~1 and~2 in Labrador \&\ Mewaldt 1997.)
Therefore, excesses of antiprotons 
observed at these low energies
may be hard to interpret as evidence for
dark matter.  For this reason, we will be interested in the flux of
interstellar antiprotons in a small window of kinetic energy, $\Delta K$,
below $1\gev$.  The observable we want to measure is
\beq
I_{\bar p}(\Delta K)=\int_{\Delta K} dK \Phi_{\bar p}^{\rm IS}(K)\epsilon (K)
\eeq
where $\epsilon (K)$ is the experimental 
acceptance efficiency of an antiproton
with kinetic energy $K$.  The units of $I_{\bar p}(\Delta K)$
are ${\rm cm}^{-2}\, {\rm sec}^{-1}\, {\rm sr}^{-1}$.
In the next section we shall discuss the experimental apparatus to
measure this with $\Delta K$ corresponding to $50\mev < K< 200\mev$.

Another potential source for interstellar
antiprotons is evaporation of Primordial Black Holes (PBHs)
(Hawking 1974; Carr 1985; Page \&\ Hawking 1976; Kiraly et al. 1981;
Turner 1982).
These can be generated if significant density fluctuations
arise in the early universe causing gravitational collapses in overdense
regions (Hawking 1971; Carr 1985).
Other ways to produce PBHs have been envisioned
(Hawking et al. 1982; Hawking et al. 1989).
Most of
the mechanisms admit scale-invariant density perturbations which leads
to a continuous spectrum of masses.  The number density can be 
parametrized (Carr 1975) as
\beq
\label{pbhspectrum}
\frac{dn}{dM}=\frac{(\beta -2)\Omega_{\rm PBH}\rho_c}{M_H^2}
\left( \frac{M_H}{M}\right)^\beta
\eeq
where $\rho_c$ is the closure critical density of the universe,
$\Omega_{\rm PBH}$ is the fraction of $\rho_c$ attributable to
PBHs which have not
yet evaporated, and $\beta$ is calculated from the equation of
state when the PBHs were formed ($\beta =2.5$ in the radiation-dominated
era).

Standard Hawking evaporation calculations show that any PBH which has
mass less than $M_H\simeq 10^{15}$~grams (the ``Hawking mass'')
would have already decayed
before our present time
(Page 1976; MacGibbon \&\ Carr 1991; Halzen et al. 1991).
The PBHs which are of most interest in
seeing evaporation products are those black holes which are currently
decaying, and so have mass near $M_H$.  Thus, limits on primordial black
hole density ($\Omega_{\rm PBH}$) come from the number density of PBHs
with $M\sim M_H$.  Using the simplest scale-invariant density perturbation
assumption which leads to Eq.~\ref{pbhspectrum} one 
finds (MacGibbon \&\ Carr 1991; Halzen et al. 1991) that
$\Omega_{\rm PBH}\lsim 10^{-8}$
from diffuse $\gamma$-ray observations.  Roughly equivalent
limits (Maki, Mitsui, \&\ Orito 1996; Mitsui, Maki, \&\ Orito 1996)
can be found
from currently available $\bar p$ measurements (Yoshimura et al. 1995;
Mitchel et al. 1996; Moiseev et al. 1997; Matsunaga et al. 1998;
Orito et al. 1998).
An interstellar antiproton
spectrometer would significantly increase these limits, and could
allow for discovery.

One might think that such apparently extreme
limits on $\Omega_{\rm PBH}$ seem to
imply that PBHs are irrelevant to the structure and history of
the universe. However,
it should be noted, for one, that if the evaporated PBHs leave behind 
stable objects with mass above $m_{\rm PL}$,
they could provide the
cold dark matter of the universe even if $\Omega_{\rm PBH}< 10^{-8}$
(MacGibbon 1987). Therefore, PBHs with mass density
near the current limit is an interesting possibility with potentially
dramatic dark matter considerations~\fnote{a}{This should not be confused
with near critical mass density PBHs with a mass spectrum severely
peaked at $\sim 1M_\odot$ from non-scale-invariant 
perturbations (See, for example, Kawasaki, Sugiyama, \&\ Yanagida 1997).}.

The $\bar p$ profile of PBH evaporation is slightly
different than that of LSP annihilations
and so a detailed $\bar p$
energy spectrum might be able to 
resolve the differences if hints of new physics
beyond spallation appear to be required to explain data.  However,
like LSP annihilations, the large excesses of $\bar p$'s are expected
to be in the sub-GeV region.  Therefore, the
crushing effects of solar modulation muddy the experiments performed
in the heliosphere near the earth.

The solution to these problems is to not be affected by solar modulation.
Any experiment near the earth will necessarily be affected by the
solar wind even at solar 
minimum (Perko 1987; Webber \&\ Potgieter 1989; Gaisser \&\ Schaefer 1992;
Labrador \&\ Mewaldt 1997; Simon, Molnar, \&\ Roesler 1998).
Therefore, to more effectively probe
the antiproton spectrum for signs of exotic processes, such as LSP annihilation
and PBH evaporation, it would help to put an antiproton spectrometer
in interstellar space beyond the heliosphere, 
and therefore beyond the reach of significant solar modulation
effects.  Such a spectrometer could be one component of the payload
in the recently discussed interstellar probe effort
(Mewaldt et al. 1994).

In Fig.~1 we show the signal expected from the annihilations of
the supersymmetric LSP, and also the decay of a PBH.  The expectations
have several astrophysical uncertainties.  The most important uncertainty
is the density profile of dark matter in our galaxy.  The LSP curve
shown in Fig.~1 is for the supersymmetric model of
Bottino et al. 1998 with $m_{\rm LSP}=62\gev$, the LSP is mostly a gaugino-like
neutralino with less than 2\% Higgsino component, 
and $\Omega_{\rm LSP}h^2=0.11$. The assumed density profile is
\beq
\rho_{\rm LSP}(r,z)=\rho_{\rm loc}\left\{ 
 \frac{a^2+r^2_{\odot}}{a^2+r^2+z^2}\right\}
\eeq
where $a=3.5\, {\rm kpc}$, $r$ is the radial coordinate from galactic
center, and $z$ is the coordinate perpendicular to the galactic plane.
Recent simulations (Navarro, Frenk, \&\ White 1996)
indicate that the density profile may be much more
cusping than the one given above.  This cusping possibility introduces
a systematic uncertainty in the astrophysics modeling of the dark-matter
distribution.  Although cusping would increase the antiproton signal, other
uncertainties, such as clumping of dark matter in the halo, may decrease
(or perhaps increase) the antiproton signature.
The uncertainties in not knowing the precise
particle physics model, and not knowing the precise astrophysical model
of dark matter make it difficult to make definitive predictions of
the signal antiproton flux.  However, one does generically expect the
flux of $\bar p$'s below $1\gev$ due to near critical density supersymmetric
dark matter to be comparable to secondary $\bar p$'s.

In Fig.~1 we have also plotted the expected interstellar antiproton flux
from a PBH evaporation calculation of Maki, Mitsui and Orito 1996.
This is not the most optimistic prediction, nor is it the
most pessimistic prediction of PBH evaporation.  It is a prediction
based on a reasonable set of assumptions and parameters. An upward turn
or enhancement of the antiproton spectrum for kinetic energies below
$1\gev$ would be an impressive clue that
non-standard physics processes are occuring in the galaxy.

\section{Experimental discussion}

           Very limited weight and power will be available for any
 experiment on board an interstellar probe. Those constraints dictate
 the design. We propose to use the annihilation signature of
 antiprotons that stop in a block of heavy material and release their
 entire rest mass energy ($\sim 938\mev$). We base our design on a
 cube of heavy scintillator (BGO) with mass of the order of 1.5~kg. 
That cube, $42\, {\rm g/cm}^2$ thick, will stop antiprotons (and protons)
 of energy $\lsim 250\mev$. A time-of-flight system (TOF) is used to
 select low energy, slow particles. Particles with energy less than 
$\sim 50\mev$ will not penetrate to the main crystal through the TOF
 counters, so this sets the lower energy limit.

            The separation of antiprotons from protons is the most
 challenging aspect of the design. Any low energy ($< 250\mev$) proton
 which would pass TOF selections cannot deposit more than its own
 kinetic energy in the block while stopping antiprotons release their
 rest mass energy through the annihilation. Antiprotons will be
 required to deposit more than 300~MeV. A proton can deposit
 comparable energy in this amount of material only through hadronic
 interaction; only protons with energy $\gsim 500\mev$ can
 efficiently do so. These higher energy protons will be separated from
 $< 250\mev$ antiprotons by the TOF.  As a conservative estimate we
 assume that all protons with energy above 500~MeV have the potential
 to create a background of ``antiproton like'' events, and their
 integral flux in interstellar space would be 
  $ \sim 1\, {\rm cm}^{-2}\, {\rm s}^{-1}\, {\rm sr}^{-1}$ 
 (using local interstellar proton flux from Seo et
 al. 1987). An example antiproton flux from PBH evaporation
 is $\sim 10^{-6}\, {\rm cm}^{-2}\,{\rm s}^{-1}\, {\rm sr}^{-1}$ 
 in the energy
 interval from 50 to 200~MeV (Maki, Mitsui and Orito, 1996); the
 expected secondary antiproton flux is a factor of ten lower (see
 Fig.1). This gives us the requirement to have no more than one false
 ``antiproton'' from $10^7$ protons, an extremely challenging task.

             The proposed instrument is shown in Fig.~2. We plan to use
 a $6\, {\rm cm} \times 6\, {\rm cm} \times 6\, {\rm cm}$ 
segmented piece of a heavy
 scintillator such as BGO to stop antiprotons and measure the energy
 of their annihilation. The choice of $\sim 250\mev$ as the highest
 antiproton energy to be detected is a compromise between desirable
 higher detectable energy and better time-of-flight separation.  This
 energy brings us to the chosen dimensions of the crystal keeping in
 mind the necessity to minimize the detector's weight. 
 Segmentation of this block is
 needed to remove high-Z low-energy nuclei, which deposit energy in a
 predictable, continuous pattern.  It also helps to remove protons with
 energy between 300~MeV (energy threshold in the crystal) and 500~MeV,
 which have large scattering in the crystal and consequently a longer
 path and larger energy deposition. Making the stopping block a
 $6\times 6\times 6$ array of $1\, {\rm cm}^3$ crystals will also allow us to
 form a crude image of the pattern of energy deposition. The TOF system,
 consisting of four 5~mm thick plastic scintillators spaced by 5~cm,
 selects only low energy slow particles. The scintillator closest to
 the central BGO is $6\, {\rm cm} \times 6\, {\rm cm}$ 
and is divided in 2 strips, while
 the outer scintillator is larger, $12\, {\rm cm} \times 12\, {\rm cm}$ with 3
 strips. Two scintillators in the middle of the TOF stack are $10{\rm cm}
 \times 10\, {\rm cm}$ and $8\, {\rm cm} \times 8\, {\rm cm}$.  
  The trigger is the coincidence
 of all possible pair combinations (6) of time-of-flight detectors and
 must be above a time-duration threshold which corresponds to $\beta
 \simeq 0.7$.  Moreover, all pulse heights from the 4 scintillators should be
 above a threshold which corresponds to the ionization loss for the
 appropriate velocity particle. This threshold will be about twice the
 mean energy loss of a minimum ionizing Z=1 particle (\it mip) \rm to
 find slow particles and reject faster (lower \it dE/dx) \rm
 ones. Finally, the trigger will require that the energy detected by
 the BGO crystal be above $\sim 300\mev$.  A fast proton (\it mip) \rm
 without an interaction loses $\sim 90\mev$ in the worst case of
 longest path traversed 
 in the crystal. The time resolution of TOFs is assumed to
 be $50\, {\rm ps}$ which is probably the best that can be presently achieved
 with scintillators of this size (Mitchell, private
 communication). All possible ways to reach and improve on this
 resolution should be explored.  Currently we have simulation results
 that indicate a proton rejection power of $2\times 10^6$ with a good
 hope of reaching the requirement by tightening selections and finetuning 
 the instrument design. For example, additional rejection power
 can be obtained using more sophisticated on-board selection of events
 which corrects \it dE/dx \rm and transit time measurements from the
 scintillators for arrival direction.  A major source of background in
 our simulations to date is due to timing fluctuations; chance
 coincidences will also play a role and must be explored carefully. We
 can also compare the pattern of energy deposition in the individual
 crystals to that expected from annihilations and interactions to
 achieve further background rejection.  The efficiency, $\epsilon (K)$, of the
 antiproton acceptance after all the selections described above are
 applied is shown in Fig. 3. The energy resolution is $\sim 10\%$ at
 100~MeV, provided by TOF. The estimated dimensions of the instrument
 are $ 25\, {\rm cm} \times 20\, {\rm cm} \times 20\, {\rm cm}$ 
 with $5\, {\rm kg}$ weight and $15\, {\rm W}$ of
 power. Experimental mass, the most critical parameter for a deep-space
 mission, is broken down into BGO crystal ($2\, {\rm kg}$), the TOF
 scintillators with frame and phototubes ($1.6\, {\rm kg}$), 
  electronics ($1\, {\rm kg}$) and mechanical structure 
 ($0.4\, {\rm kg}$).  An on-board processor would
 analyze the data and reject most residual background.  The data rate
 would be very low, $\sim 1000\, {\rm bit/day}$.

             The estimated geometrical factor for this instrument
 would be $G\simeq 10 \: {\rm cm}^2\, {\rm sr}$. 
The expected event rate would be 0.3 -
 3 antiprotons per day between 50 and 200~MeV energy.  We can estimate
the exposure time, $t_e$, needed to measure a signal to within
$x\%$ statistical uncertainty:
\beq
t_e =\frac{100}{I_{\bar p}(\Delta K)}\frac{1}{G}
 \left( \frac{10\%}{x}\right)^2 .
\eeq
Therefore, to obtain a $10\%$ statistical precision measurement of the
example $\bar p$ flux from PBH evaporation given in Fig.~1, where
$I_{\bar p}(\Delta K)\simeq 4\times 10^{-7}\, {\rm s}^{-1}\, {\rm sr}^{-1}\,
 {\rm cm}^{-2}$, one would require approximately 
$t_e\simeq 0.8\, {\rm yr}$ of exposure
time. Based on statistical inference only, both the LSP signal and
the PBH signal in Fig.~1 would be detectable in approximately a year.
Since exposure times for many years are possible for an interstellar
mission, one could reduce the bandwidth and efficiency for more rejection power
against protons. 

\section{Conclusion}

The quest for dark matter has been a long one.  Despite recognizing
the potential existence of weakly interacting stable particles for
many years, we have yet to find it or rule it out.  The challenge
is most explicitly made clear in a supersymmetric context, since a large
class of models yield an LSP relic abundance near critical
density (Wells 1998).  Within these particular
models, past and present
dark-matter experiments are likely not to have a signal for LSP annihilations
unless 
optimistic astrophysical parameters are assumed. Future 
detectors will cover much ground in the supersymmetric parameter
space.  As outlined in the beginning, each experiment
has its advantages and weaknesses even within the specific supersymmetric
framework.  The advantage of the interstellar
antiproton search is that it is more
insensitive to the type of dark matter (WIMPs, LSPs or PBHs), and the signal is
always expected to be continuous against a peaking background.
Furthermore, measurements of the interstellar $\bar p$ flux would
greatly aid the analysis of data from other experiments taking
place near the earth (AMS, PAMELA, and others), since the interstellar $\bar p$
flux would be an extremely useful observable
to compare with the inverse modulated near-earth data.
Exploring how the spectrum changes as the detector leaves the heliosphere
would be an added bonus to this project.  This could further help our
understanding of energy losses and flux modulations in the solar system.
It is mainly for these
reasons that we think an interstellar antiproton detector would be
an excellent addition to the search for dark matter.

\noindent
{\it Acknowledgements:}  One of us, JFO, would like especially to thank 
the management at the
Stanford Linear Accelerator Center for their hospitality during the period
in which this work was initiated.


\clearpage
\centerline{\bf Figure Captions}

\medskip
\noindent
Fig.~1.--
{\protect The 
$\bar p$ interstellar flux, $\Phi_{\bar p}^{\rm IS}(K)$, versus kinetic
energy from ordinary spallation processes
(solid line --- Simon, Molnar, \&\ Roesler 1998), from
supersymmetric LSP annihilations for a particular set of
supersymmertry breaking parameters with $m_\chi =62\gev$
(dashed line --- Bottino et al. 1998), 
and primordial black hole evaporation (dash-dotted line --- Maki, Mitsui,
\&\ Orito 1996).}

\medskip
\noindent
Fig.~2.--
{\protect Diagram of the experimental design. See text for details.}

\medskip
\noindent
Fig.~3.--
{\protect The efficiency $\epsilon(K)$ for antiproton acceptance as a function
of its kinetic energy after all the selections
described in the text are applied.}

\clearpage
\pagestyle{empty}

\begin{figure}[t]
\centerline{\epsfxsize=5.0truein \epsffile{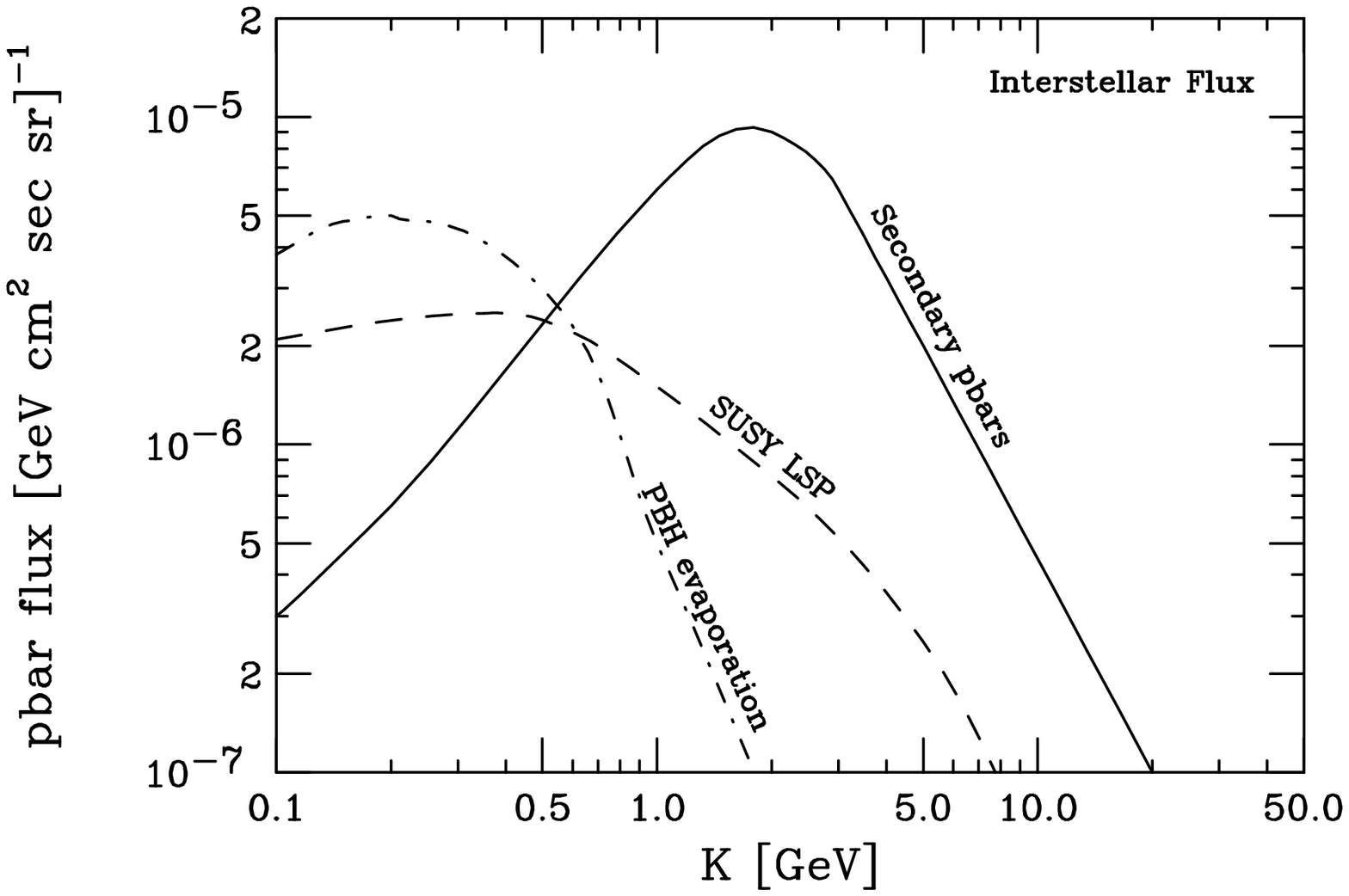}}
\bigskip
\bigskip
\hspace*{8cm}
\mbox{\Large \bf Fig. 1}
\end{figure}


\clearpage
\pagestyle{empty}

\begin{figure}[t]
\centerline{\epsfxsize=5.0truein \epsffile{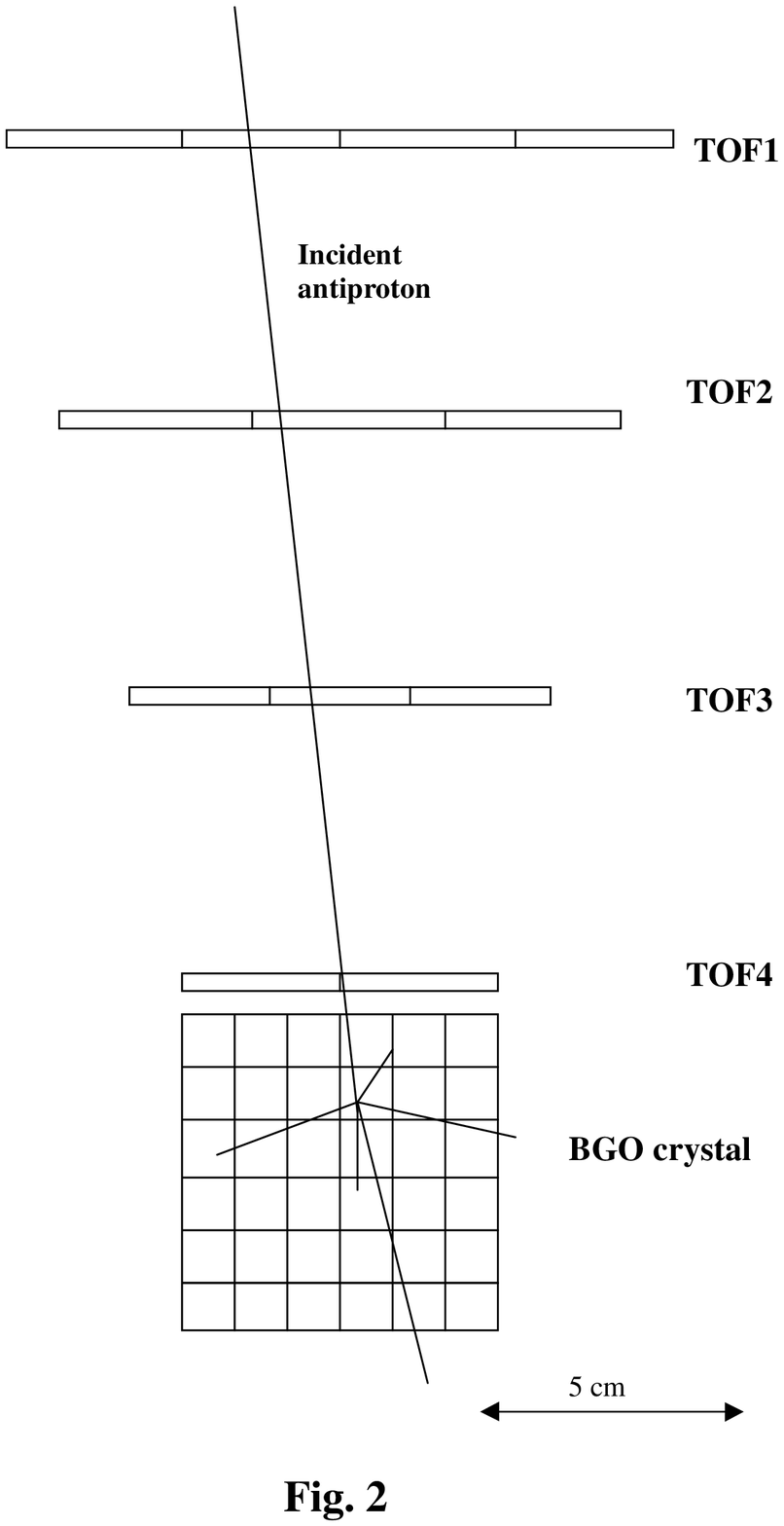}}
\end{figure}

\clearpage
\pagestyle{empty}

\begin{figure}[t]
\centerline{\epsfxsize=5.0truein \epsffile{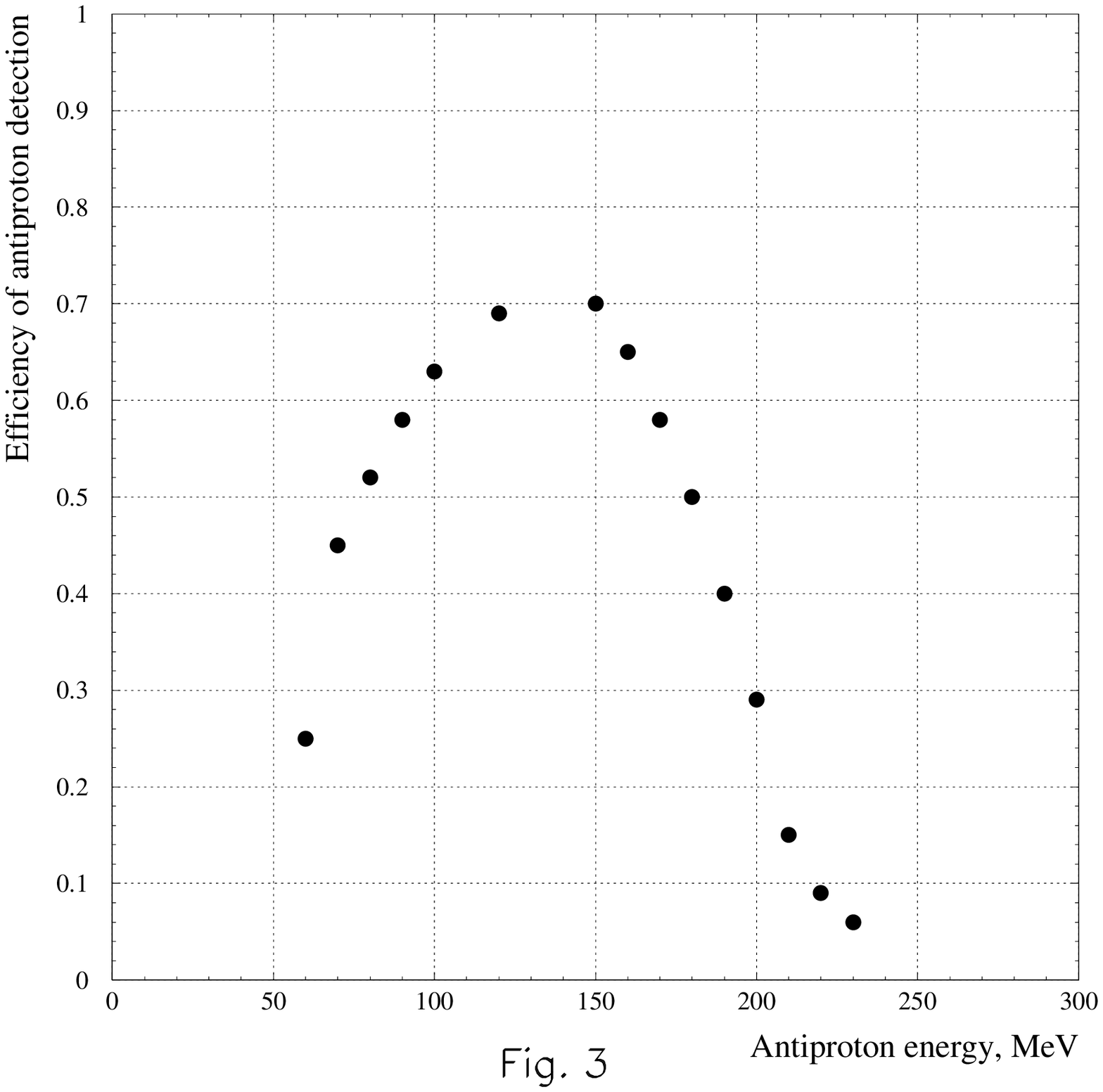}}
\end{figure}

\end{document}